# Timing-injection locking in a self-starting Mamyshev oscillator induced by the dissipative Faraday instability

**Changqing Li[1], Ran Xia[2, *], Yutai Zhao[1], Yifang Li[1], Jia Liu[1], Christophe Finot[3], Xiahui Tang[1] and Gang Xu[1, *]**

[1] School of Optical and Electronic information, Huazhong University of Science and Technology, Wuhan, China
[2] Department of Electronic Engineering, Xiamen University, Xiamen, China
[3] Université Bourgogne Europe, CNRS, Laboratoire Interdisciplinaire Carnot de Bourgogne ICB UMR 6303, 21000 Dijon, France
Email : ranxia@xmu.edu.cn, gang_xu@hust.edu.cn

### Abstract

Mamyshev oscillators (MOs), a novel class of passively mode-locked fiber lasers, serve as an excellent platform to explore complex nonlinear dynamics, ranging from localized structures to chaos. Despite their versatility, achieving self-starting mode-locking remains a significant challenge. In this study, we unveil the critical role of the dissipative Faraday instability (DFI) in facilitating the self-starting process of MOs, where the DFI triggers the symmetry breaking of the homogeneous solution to overcome the initiation barriers. A panoramic view of several distinct operational regimes with distinct DFI patterns is provided, namely the non-self-starting states, the irregular patterns, the harmonic mode locking regime, the stable single pulse and the stable multi pulse regime. For the lattest case, we uncover the origins of randomness in these pulse sequences through analyzing the causality between the timing of the random pulses and the initial seeding conditions. Building upon these findings, we propose the novel time-injection locking technique to customize the temporal locations of the pulses as well as the pattern timing in MOs, thus demonstrating its potential for applications in all-optical data storage and tunable ultrashort pulse sources.

## 1. Introduction

Passively mode-locked fiber lasers have been widely recognized as a reliable and highly competitive source of light for the generation of ultrafast pulses owing to the advantages of compactness, high beam quality and low cost.[1] These laser architectures have been employed in numerous fields such as optical telecommunication, material processing and biomedical imaging, and also provide an excellent platform for exploring complex nonlinear dynamics, including various stationary dissipative solitons,[2–4] breathing or pulsating solitons,[5,6] chaos[7] and so on. Diverse mechanisms have been explored to achieve passive mode locking, such as the nonlinear optical response of material-based saturable absorbers (SAs),[8] and the artificial SAs such as nonlinear polarization evolution[9] and nonlinear amplifying loop mirror.[10] More recently, Mamyshev oscillator (MO), a novel scheme of passively mode-locked laser system, has been attracting intensive attention[11–13] mainly due to the ability of generating ultrashort Megawatt pulses.[14,15] The concept of MOs derives from the optical Mamyshev regenerators (MRs) used in the cavity as an effective SA.[16,17] The MR initially developed for high speed telecommunications applications[11,18] relies on the spectral broadening induced by self-phase modulation (SPM) in a normally dispersive fiber, followed by offset spectral filtering, leading to a nonlinear optical power response similar to SA, associated with an optical limiting and pulse reshaping abilities. MOs consist of a pair of frequency-detuned MRs in a loop cavity, which enables the generation of well-defined pulses, the nonlinear distortions accumulated over one cavity roundtrip being canceled by the MRs. In general, the concatenation of the transmission functions of Mamyshev SA approximates a step-like function with large modulation. [17] If the main application of MOs has been up-to-now the generation of high-energy ultrashort pulses, it should also be noted that many nonlinear dynamics observed in more traditional fiber architectures have also been reported in MOs, which is now a part of the large family of highly dissipative soliton structures. About these different characteristics, we can mention the harmonic mode-locking operation [19–21], the subharmonic operation with the presence of breathing structures and route to chaos [22–24] and the formation of bound pairs [25,26].

Among the challenges for the MO is its non-self-starting nature. Indeed, due to the cancelation of the CW components by the offset filters, the mode-locking initiation from noise remains an intrinsic challenge for MOs. Even though the parametric gain of the modulation instability (MI) may lead to the growth of the high-power patterns to initiate the mode-locking process, the phase matching conditions of the conventional MI, including the Benjamin Feir MI [27-29] and the Turing MI [30] are unable to be naturally satisfied in the regime of normal dispersion. Some practical solutions have been proposed for high-peak-power pulse generation, [31,32] and a few studies have revealed that, under specific cavity parameters configurations, even in normal dispersion regime, the self-starting of MOs could be triggered by the dissipative Faraday instability (DFI) [33–36]. As a novel type of MI, the DFI could be originated from the periodic modulation of spectral losses in MOs due to the offset filters, thus spontaneously breaking up the CW and giving birth to stationary solitons. [33] In such normally dispersive MOs, the DFI is a golden key to initiate mode locking from background noise and subsequently stabilize the periodic pulse train without the requirement of external seeding or specific cavity manipulation. To date, high-repetition-rate harmonic mode locking driven by the DFI has been experimentally demonstrated in a 1.5-µm Raman laser [35] and theoretically explored in rare-earth doped fiber lasers at 1-µm [33,37] and 2-µm wavebands [38]. It is worth emphasizing that, contrary to the traditional harmonic or multi-pulse mode-locking that typically involve pump powers higher than the one required for single pulse operation [19–21], the DFI-assisted mode locking results from a very different nonlinear dynamics and can generate high repetition pulse train at relatively low pump powers. However, for given driving parameters in this harmonic mode-locking regime (HR), the temporal output of MOs will be filled with equally-distant pulses, each peak having a fixed spot. Therefore, the manipulation on the pulse train, not only for intensity but also for pulse timing becomes challenging. Here, unlike previous investigations on the DFI, we focus more on the DFI-induced generation of well-defined pulses at one or more random temporal locations. We discuss the controllability that can be obtained over this process and how it can lead to the the basis for all-optical storage and buffering.

In this work, we therefore conduct comprehensive numerical and theoretical investigations into the panoramic view of operational regimes and the self-starting mechanism of random pulse train in an all-normal-dispersion MO. Based on the nonlinear Schrödinger equation (NLSE), we calculate the parametric gain of the DFI using the linear Floquet stability analysis. The dependence of the gain spectrum on the pumping power and wavelength detuning facilitates classification of various operation regime of MOs, including the irregular operation regime (IR), the harmonic operation regime (HR), the non-self-starting operation regime (NR) and the random operation regime (RR). Along this line, we present the causality between the timing of pulses in the RR and the initial weak seed condition. Consequently, we discuss a novel scheme for programmable generation and storage of soliton sequences by external injected pulse with high peak power. The timing of the generated soliton train synchronizes or 'locks' with the injected signal. This phenomenon resembles the injection locking, in which the oscillator frequency and phase are locked to those of the injected pulses [39]. Accordingly, we term the new phenomenon *timing-injection locking* (*TIL*).

## 2. Analysis of the self-starting pulse generation process

### 2.1. Propagation equations and simulation model

The schematic diagram of a ring MO is shown in Fig. 1(a). The unidirectional cavity is composed of two arms, each consisting of a section of Ytterbium (Yb)-doped gain fiber (YDF), a section of single mode fiber (SMF) and an optical band pass filter (OBPF). Within the cavity, the pulse spectrum undergoes broadening in the YDFs and SMFs and is subsequently shaped by the OBPFs. To simplify the model, the only distinction between the two arms is set to be the central wavelength of the OBPFs. As a first approximation, the propagation of pulses could be described by the nonlinear Schrödinger equation (NLSE) including energy and frequency dependent gain, which could be written as [33, 40]:

$$\frac{\partial A}{\partial z} = -\frac{\alpha}{2}A - \frac{i\beta_2}{2}\frac{\partial^2 A}{\partial t^2} + i\gamma|A|^2A + \frac{g}{2}A + \frac{g}{2\Omega_g^2}\frac{\partial^2 A}{\partial t^2}.$$

In this equation, $A$ is the slowly varying envelope of the electric field. $z$ is the propagation distance and $t$ is the time in a reference frame moving with the group velocity. $\beta_2$ is the second-order dispersion coefficient and $\gamma$ is the nonlinear coefficient. $g$ is the saturable gain coefficient and is zero for the SMFs. The gain coefficient could be described as $g = g_0/(1 + \frac{E_{pulse}}{E_{sat}})$, where $g_0$ is the small-signal gain, $E_{pulse}$ is the pulse energy, and $E_{sat}$ is the saturation energy. $\Omega_g$ is related to the gain bandwidth of the YDFs. Here, it is worth mentioning that the intra-pulse Raman contribution is not considered as, in this normally dispersive cavity, the pulses are less affected by self-frequency shift, while it could be a crucial element of the dynamics in MOs operating in anomalous dispersion. [41] Third order dispersion is also a factor we neglect.

The spectral transmission function of the OBPFs is set to be a fourth-order super-Gaussian shape and is expressed as: $F_{\pm}(\omega) = \exp[-\frac{(\omega \pm \delta\omega)^4}{\sigma^4}]$ with $\delta\omega = -2\pi c\frac{\delta\lambda}{\lambda_0^2}$ and $\sigma = \pi c\frac{\Delta\lambda}{(ln2)^{1/4}\lambda_0^2}$. Here, $\delta\omega$ is the central frequency detuning, related to the central wavelength detuning $\delta\lambda$ of the OBPFs (see Fig. 1). c = $3\times10^8$ m/s is the light speed. $\lambda_0$ is the central wavelength of the laser system. $\Delta\lambda$ is the 3-dB spectral bandwidth of the OBPFs. Note that, $\lambda_0$ = 1060 nm and $\Delta\lambda$ = 3 nm are fixed in our simulations, while the wavelength detuning $\delta\lambda$ is used to describe the symmetry offset of the central wavelength of the OBPFs compared to $\lambda_0$.

The following parameters are used in accordance of the experimental devices [22,23]: $g_0$=10 m$^{-1}$, $\Omega_g$= 50 nm, $L_{YDF1} = L_{YDF2}$ = 1 m, $\beta_{2\text{-}YDF}$=30 ps$^2$/km, $L_{SMF1} = L_{SMF2}$ = 60 m, $\beta_{2-SMF}$= 20 ps$^2$/km, $\gamma_{YDF} = \gamma_{SMF}$ = 3 W$^{-1}$km$^{-1}$. The output coupling ratio of the fiber coupler is 0.1. With such settings involving long segments of fibers, the goal of our cavity is not to generate the high-peak power pulses but to generate more pulses over one cavity roundtrip. Numerical simulations are performed using the standard split-step Fourier method [40], considering a temporal window of 400 ps in order to maintain a reasonable computation time. The degrees of freedom of the MO are numerous, and in this contribution, we mainly focused on the impact of the saturation energy $E_{sat}$ and the wavelength detuning $\delta\lambda$ to describe the pump energy and the separation of the OBPFs, respectively. We will more briefly discuss the influence of the length of the SMF segments.

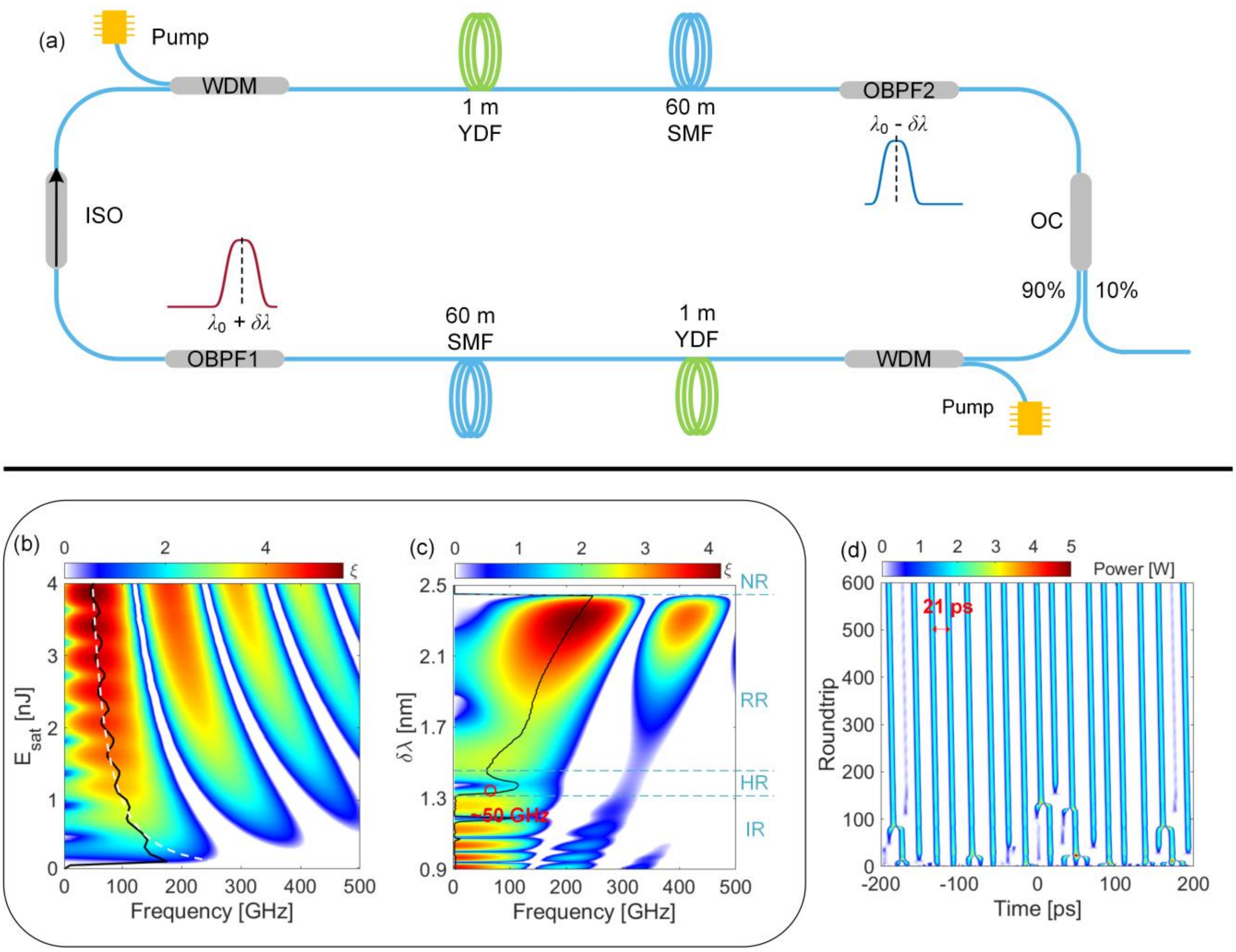

**Fig. 1. Schematic diagram and mechanisms of the self-starting MO.** (a) Pump: LD pump laser at 980 nm; WDM: wavelength division multiplexer; YDF: Yb-doped gain fiber (in green, normal dispersion); SMF: single-mode fiber (in black, normal dispersion); ISO: optical isolator; OBPF1: longer-wavelength super-Gaussian spectral filter; OBPF2: shorter-wavelength super-Gaussian spectral filter. The wavelength detuning $\delta\lambda$ is used to control the spectral offset of the two OBPFs. (b) Influence of the pump parameter, $E_{sat}$ on the instability gain spectrum of the DFI. The parametric resonance tongues of the DFI are obtained by the linear Floquet stability analysis, which is performed by calculating the growth exponent of the perturbation over one cavity roundtrip. The black line shows the frequency of the perturbation with the maximum instability gain $f_{\text{max-gain}}$ at each saturation energy $E_{sat}$, whereas the dashed white line shows the trend $f_{max-gain} \propto E_{sat}^{-1/2}$. The detuning parameter, i.e. the wavelength offset of the two filters, is set to be $\delta\lambda$ = 1.8 nm. (c) Influence of the detuning parameter, $\delta\lambda$ on the instability gain spectrum of the DFI. The black line shows the frequency of the perturbation with the maximum instability gain $f_{\text{max-gain}}$ within the first instability sideband at each detuning $\delta\lambda$. The pump parameter is $E_{sat}$ = 0.5 nJ. The horizontal dashed lines represent the borders of the various regimes of operation of the MO. (d) Temporal evolution of the harmonic mode locking. The time interval of the multipulses corresponds to the repetition rate obtained from the Floquet analysis. The key parameters are $E_{sat}$ = 0.5 nJ and $\delta\lambda$ = 1.35 nm [see red circle on panel 1(c)].

In this MO structure, the intra-cavity light field undergoes the periodic modulation of spectrally dependent loss caused by the detuned OBPFs. To elucidate the onset of DFI patterns, we analyzed the Faraday instability gain spectrum, using the numerical linear Floquet stability analysis [34–36], as depicted in Fig. 1(b, c). The numerical stability analysis in a periodic system was conducted by computing the evolution of perturbations of each mode over one cavity roundtrip. The results show which modes' perturbations experience amplification due to the periodic loss, i.e. the gain through loss process. [42] Here, we defined the instability gain as $g_{in} = \xi/L_t$, where $L_t$ is the total length of the cavity and $\xi$ is the average growth exponent for each mode in one cavity. $\xi$ is calculated by $\xi = \ln(F_m(\omega))$ with $F_m(\omega) = \max(|F(\omega)|)$, where $F(\omega)$ is the Floquet multiplier. Figure 1(b) summarizes the effect of $E_{sat}$ on the instability gain spectrum for a fixed detuning $\delta\lambda$ = 1.8 nm. The frequency corresponding to the maximum instability gain $f_{\text{max-gain}}$ within the first instability sideband decreases with the gain saturation energy $E_{sat}$, following a trend: $f_{max-gain} \propto E_{sat}^{-1/2}$, reported in several previous works [34–36] [see dashed white line in Fig. 1(b)].

Generally, in the Mamyshev configuration, the wavelength detuning $\delta\lambda$ plays a key role in adjusting the balance of loss and gain. Therefore, it is essential to carefully investigate the transition between different behaviors as a function of $\delta\lambda$ [37], as some studies have already shown a transition between ignition and pulse buffering behaviors of self-pulsating MRs-based sources [43] while maintaining a constant saturation energy $E_{sat}$. In that context, figure 1(c) illustrate the evolution of the instability gain spectrum with fixed $E_{sat}$ = 0.5 nJ. The black line shows the frequency of the perturbation with the maximum instability gain $f_{\text{max-gain}}$ within the first sideband at each detuning $\delta\lambda$. The first sideband appears to dictate the overall behavior of the cavity. Quite remarkably, three critical inflection points at 1.3 nm, 1.4 nm, and 2.4 nm, demarcating four distinct operational regimes in the MO can be identified. When $\delta\lambda$ < 1.3 nm, the maximum instability gain dominates at zero frequency and the pulses are unstable at this operational regime, what we call irregular operation regime (IR). Note that for some very specific values of $\delta\lambda$, the second instability gain sideband may exhibit a gain higher than the first sideband. However, we have checked that it does not qualitatively affect the nature of the operational regime. When 1.3 nm< $\delta\lambda$ < 1.4 nm, in the harmonic operation regime (HR), MO tends to generate a train of closely arranged pulses with a repetition rate corresponding to $f_{\text{max-gain}}$ obtained from the Floquet analysis [see Fig. 1(c, d)]. When 1.4 nm< $\delta\lambda$ < 2.4 nm, contrary to the strictly high-repetition-rate harmonic mode locking, the more frequent pulse collisions break the stability of the multi-pulse distribution in all-normal-dispersion MOs, leading to a random operation regime (RR) as stressed in the pioneering experimental investigations involving cascaded MRs [44]. Whereas, when changing a perspective, all-normal-dispersion MOs emerge the capacity of producing a wealth of unequally spaced pulses, providing the flexible platform to get insights into the complex pulse dynamics and further manipulation on multipulses. When $\delta\lambda$ > 2.4 nm, the MO is unable to self-start from noise due to the absence of instability gain.

## 2.2. Panoramic view of operational regimes

Achieving self-starting or sustaining a single or multiple pulse is therefore closely related to the MO parameters. [37,38,45–48] We provide in this section a comprehensive panoramic view of different operational regimes according to the two critical variables that are the wavelength detuning $\delta\lambda$ and saturation energy $E_{sat}$. Specifically, we performed a large set of systematic numerical simulations by changing the values of $E_{sat}$ with a step of 0.05 nJ and $\delta\lambda$ with a step of 0.05 nm while the other parameters are set as above and the initial condition of simulations is random noise. The output patterns achieved after 600 roundtrips were categorized according to the four previously mentioned operation regimes and results are summarized in the panoramic view shown in Fig. 2(a), with typical temporal behavior illustrated in Fig. 2(c2) and spectral filtering properties provided in Fig. 2(c3). Figure 2(a) points out that the parameter space of the various operation regimes is highly correlated with the OBPF detuning, while the value of $E_{sat}$ has less impact. The demarcation of the regimes corresponds to the inflection points obtained from the Floquet analysis as discussed in the previous section from the specific value $E_{sat}$ = 0.5 nJ. Indeed, we have checked the instability gain spectrums exhibits, for a large range of values of $E_{sat}$, similar trend and inflection points as the one reported in Fig. 1(c).

The irregular regime (IR, green background and circles) is characterized by a non-stationary pattern with the frequent emergence of pulse collisions and annihilation [see Fig. 2(c) for typical spectral detuning $\delta\lambda$ =1 nm]. Indeed, the weak wavelength detuning induces a significant overlap of the OBPF tails and fails to suppress the continuously growing background noise between self-started pulses. A typical phenomenon, the emergence of optical rogue waves, is also observed in this pattern during the nonlinear collision process as discussed in prior work. [36] The yellow background represents harmonic mode locking regime (HR) where the typical period of the pulse train is directly fixed by the frequency of the maximum of the instability gain spectrum. Note that the existence of this regime is in part affected by the value of $E_{sat}$.

The red pentagrams represent stable pulse trains with random time intervals and number of pulses. For reduced energy saturation values, only a single stable pulse may exist in the cavity, as marked by the blue square in the panoramic view. Both cases sustaining multiple or single stationary pulses are termed as random operation regime (RR) with purple background. In that regime, no collision process occurs, preventing the emergence of rogue events. The number of pulses that may spontaneously emerge in this regime is affected not only by the wavelength detuning $\delta\lambda$, but also by the saturation energy $E_{sat}$, as illustrated in Fig. 2(b) that is built considering the average number of pulses over 5 simulations initiated with different seeds. In general, the larger pulse number relies on moderate $E_{sat}$ and $\delta\lambda$, [46] implying the gain and loss determined stable multi-pulse state in the MO. In other words, the larger the difference between gain and loss, the more stable pulses could spontaneously exist in the cavity. In contrast to previous studies on short-cavity MOs [22,23], where increasing pump power leads to giant pulses or pulsating solitons, our long-cavity MO are more likely to generate more pulses from DFI-induced initial pulse structures. Involving higher spectral detuning promotes the existence of a single stable pulse, which peak power is significantly enhanced, enabling larger SPM-induced broadening. This explains the experimental success of self-starting strategies based on progressively increasing the spectral filter separation to initiate high-peak power single pulse operation. [37,44,49]

When $\delta\lambda > 2.4$ nm (non-self-starting operation regime NR, illustrated by the black cross within the gray background), the very reduced spectral overlap between the two OBPFs induces significant intracavity losses and prevents to initiate the pulse self-starting from noise. Typically, in that case, the transfer function of the Mamyshev device exhibits an efficient rejection of the spontaneous noise, in other words behaves as a strong saturable absorber that prevents the spontaneous emergence of a pulsed structure from the DFI process.

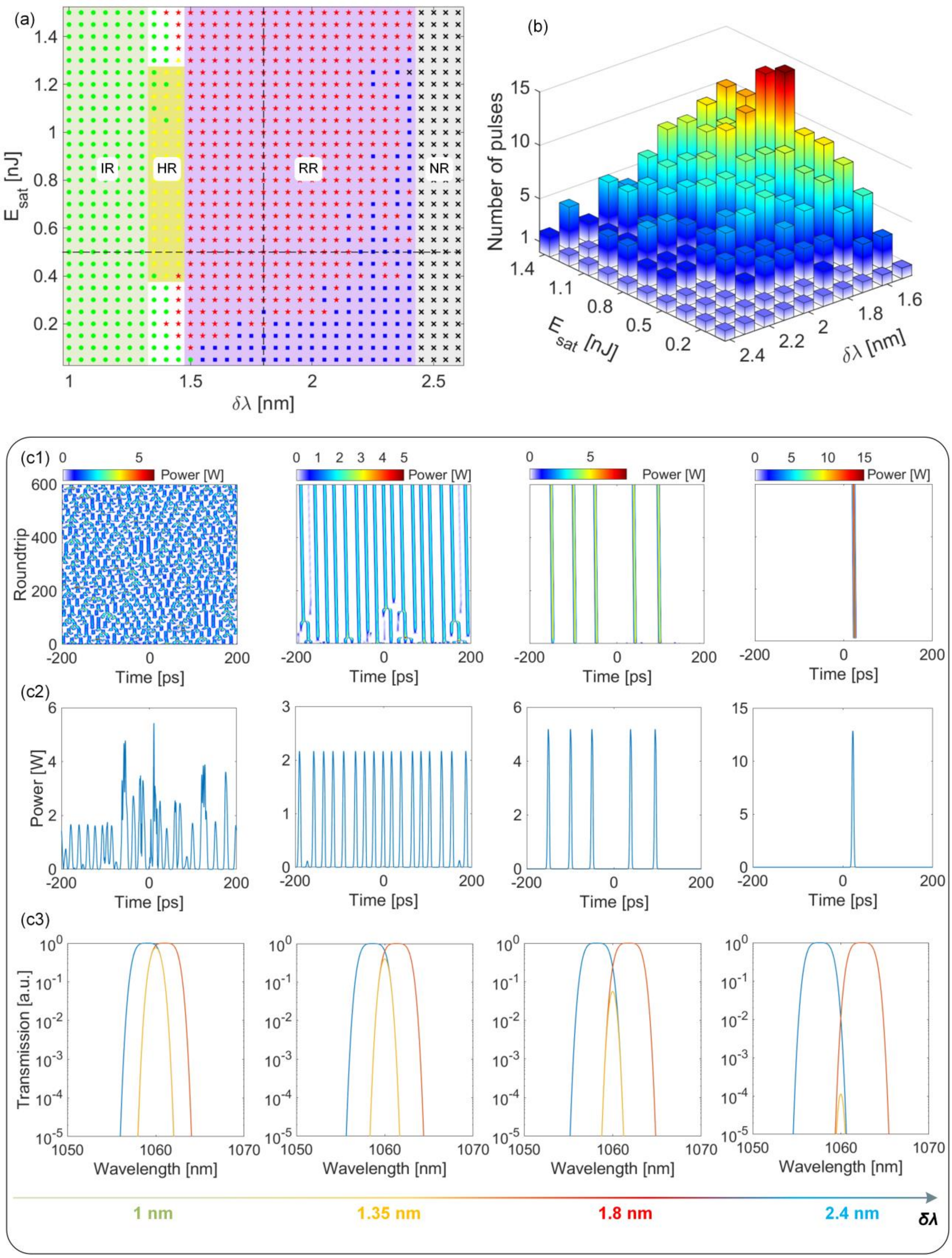

**Fig. 2. Pattern formation and self-starting dynamics in a MO mediated with the DFI.** (a) Panoramic view for operational regimes. The green circles, yellow triangle, red pentagram, blue square, and black cross represent irregular pulse train, stable pulse train, single pulse (soliton), unable to self-start (no pulse), respectively. Based on whether stable pulse or pulse train there

is in the cavity, combined with the nature of the stable pulse train, the operational regimes could be divided into irregular operation regime (IR), harmonic operation regime (HR), random operation regime (RR), and non-self-starting operation regime (NR). The four operational regimes correspond to the green, yellow, purple, and gray backgrounds in the diagram, respectively. The horizontal dashed line at $E_{sat}$ = 0.5 nJ corresponds to the case studied in Fig. 1(c) and 2(c) whereas the vertical dashed line corresponds to the case of Fig. 1(b). (b) Number of pulses emerging from noise in the RR versus $E_{sat}$ and $\delta\lambda$ for a temporal window analysis of 400 ps. Note that in the HR, the number of pulses is typically of 20 in the window. As the number of pulses may differ by one or two units depending on the initial seed under use, we have considered here the average value. (c) Typical pattern formation from noise for various operational regimes corresponding to the offsets of 1 nm, 1.35 nm, 1.8 nm and 2.4 nm. The key parameter is $E_{sat}$ = 0.5 nJ. The longitudinal evolution of the temporal intensity pattern is provided in subpanel 1, whereas the state obtained after 600 roundtrips is displayed in panel 2 and the spectral profiles of OBPF1 and OBPF2 as well as the product of the two are provided in panel 3 (red, blue and yellow lines, respectively).

Cooperated with the low wavelength detuning from our design, the self-starting phenomena can be observed in our simulations even at significant low saturation energy levels (e.g. $E_{sat}$ = 0.05 nJ). From the perspective of mode-locking mechanism, the key to initiate mode locking in a MO lies on whether the SPM-broadened spectral width is sufficient to ensure that the losses induced by the detuned filters can be compensated by the gain fiber. Therefore, the impact on self-starting under extremely low pump conditions is investigated with different wavelength detuning and lengths of the SMF, as depicted in Fig. 3(a). The initial condition for the iteration is here a Gaussian pulse seed with an extremely low peak power, ensuring that the Gaussian pulse serves as a substitute for random noise. The results based on the numerical simulations show that the larger $\delta\lambda$, the longer SMF is needed to achieve self-starting, as well as the quicker initiation of the mode locking. For instance, shot-to-shot spectral profiles before OBPF2 with 20-m or 30-m SMF in each arm under the settings of $E_{sat}$ = 0.05 nJ and $\delta\lambda$ = 2.4 nm demonstrate the failure and success of the self-starting respectively, pictured in Fig. 3(b). Specifically, both spectra experience sidebands induced by the DFI, whereas the longer fiber provides greater spectral broadening to meet the necessary spectral width for mode locking. Indeed, a longer fiber offers a larger accumulation of SPM so that the OBPFs bandwidth can be fully filled.

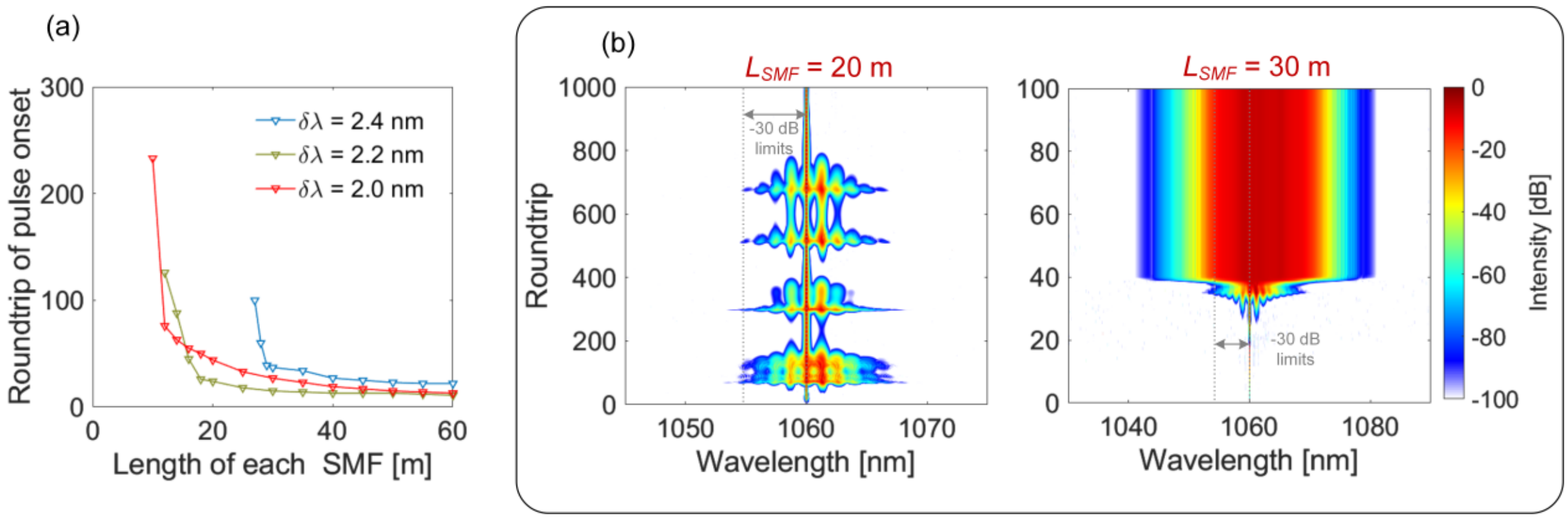

**FIG. 3. Self-starting with low pumping.** (a) Relationship between the pulse emergence roundtrips and the single-mode fiber length. Spectral OBPF detunings of 2.0 nm, 2.2 nm and 2.4 nm are tested (plotted with red, green and blue colors, respectively), whereas $E_{sat}$ is fixed at 0.05 nJ. (b) Spectral evolution before OBPF2 with $L_{SMF1} = L_{SMF2}$ = 20 m (left) and with $L_{SMF1} = L_{SMF2}$ = 30 m (right). Both cases manifest as sideband generation in the optical spectrum. This observation demonstrates that DFI-mediated spectral sideband generation precedes the establishment of stable pulse formation in the system. The key parameters are $E_{sat}$ = 0.05 nJ and $\delta\lambda$ = 2.4 nm. The vertical black dotted lines are a visual guideline to stress the -30 dB limits of OBPF2.

## 2.3. Spontaneous evolution dynamics of random pulse train

As the stable pulse trains possess more degrees of freedom with random temporal positions, it is necessary to shed new light into the buildup dynamics of the random pulse train induced by the DFI. Figure 4 demonstrates the spontaneous emergence of 4 stable pulses from noise over 40 roundtrips. We emphasize that a large number of small pulses are generated due to the DFI in the early stage, in which some pulses deplete while others are amplified through gain competition. It is worth noting that the interaction between pulses may be complex. [17,50] Here, two close pulses experience fusion, followed by energy equalization, ultimately resulting in all four pulses adopting a Gaussian-like shaped profile [see Fig. 4(a)] with a ~6 ps duration that can be directly linked to the OBPF bandwidth. These pulses present a non-negligible level of chirp that increases monotonically. The temporal evolution [see Fig. 4(c)] of the stable pulse train within the MO cavity once the stationary operation is reached shows the recovery of the temporal profile and position undergoing two rounds of 'amplification-broadening-filtering', facilitated by the cooperation of YDF, SMF and the spectral filters. The normally dispersive segment of SMF segment induces a significant temporal broadening associated with a linearization of the temporal chirp. [51] The OBPFs carve into this temporal waveform: OBPF1 isolates a part of the temporal raising edge of the pulse whereas OBPF2 isolates a part of the falling edge.

Interestingly, the fusion indicates that the generated pulse train that spontaneously emerges from the DFI can also evolve to other patterns during circling in the MO cavity, which attracted our interests on the underlying mechanism of the nonlinear dynamics. Specifically, the nearest two pulses (the 3$^{rd}$ and 4$^{th}$ pulses) gradually approach each other until their fusion while the others keep their temporal positions. It is worth noting that the energy equalization occurs before the fusion process. As marked in the red square in Fig. 4(b), the stable three pulses were initially amplified while the fused two solitons attenuated, thus eventually leading to the uniform energy distribution of the stable pulse train. The transient evolution of the temporal intensity of the intra-cavity pulses from the 15$^{th}$ to 25$^{th}$ roundtrip [see Fig. 4(d)] stresses that the overlap between the two pulses gives rise to a fast beating [see the inset in Fig. 4(d)], [52] inducing higher-frequency components compared to the initial individual pulses. Consequently, the two fusing pulses would experience larger attenuation than the other pulses, resulting in the energy change of the pulse train. Eventually, the fused single pulse settles near the temporal position of the beating, as the modulation period gradually decreases.

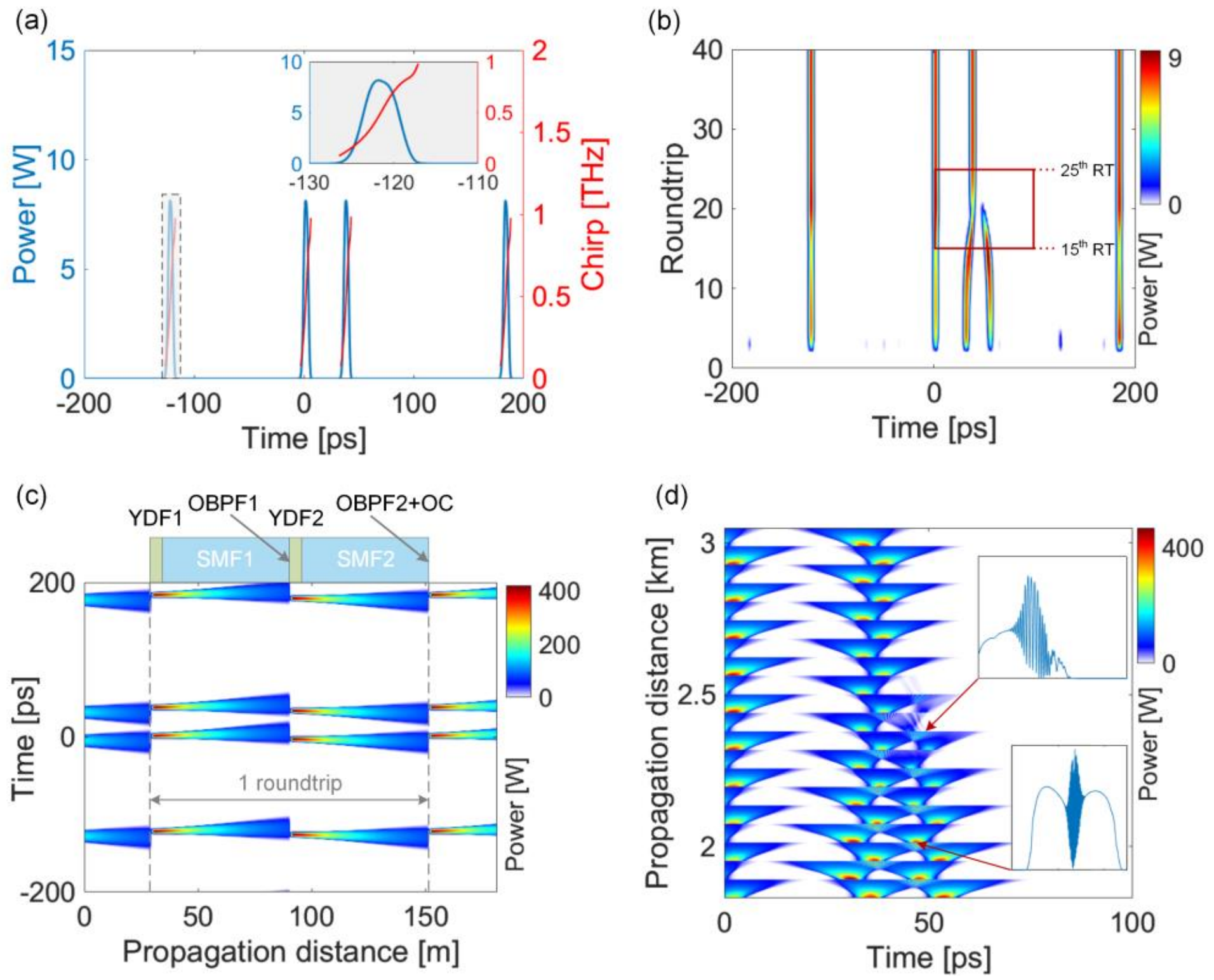

**Fig. 4. Spatiotemporal dynamics of stable pulse train.** (a) Temporal (blue) and frequency chirp (red) profiles of stable pulse train at the output port (i.e. after OBPF2). The temporal and frequency chirp profiles of a single pulse are shown in the inset with a grey background. (b) Temporal evolution over 40 roundtrips at the output port. The pulse fusion is emphasized by a red box. (c) Longitudinal evolution of the temporal intensity profile of intra-cavity pulses over one cavity roundtrip (vertical dashed lines correspond to one cavity roundtrip) when the stationary regime is reached. (d) Transient longitudinal evolution of the temporal profiles of intra-cavity pulses from the 15$^{th}$ to 25$^{th}$ roundtrip within the time window of 0 to 100 ps (see red box of panel b). Details of the temporal intensity profiles of intra-cavity pulse are plotted in the inset. The key cavity parameters are $E_{sat}$ = 0.8 nJ and $\delta\lambda$ = 1.8 nm.

## 3. Timing-injection locking and all-optical data storage in the MO

Inspired by the fusion dynamics of the DFI-induced pulse train in the MO cavity, it appears necessary to explore the causality between the initial condition and the stable pulse train pattern. Indeed, understanding the pulse buildup in MOs remains a subject of interest. [26,38,53] Here, we use seed pulses with extremely low powers, that can be seen as an embryonic light, to simulate different initial iterative conditions emerging from noise, as shown in Fig. 5. Firstly, super-Gaussian seed pulses with varying durations are investigated. We can address that the combination of dispersion, self-phase modulation, gain and optical filtering results in an output pulse distribution that is strongly affected by the temporal width of the weak super-Gaussian seed. Indeed, Figure 5(a) stresses that both the number and temporal locations of the stationary pulses are affected by the seed properties. Three evolutionary scenarios are distinguished as the single-pulse region, the fusion region and the dual-pulse region. In the first case (for temporal duration below 40 ps), as illustrated in panel (b1) for a duration of 38 ps, only one pulse emerges near the falling edge of seed pulse [see Fig. 5(b1)]. When the temporal duration is around 40 ps, two pulses can also be generated in the first stages of evolution, yet the fusion dynamic occurs and leads to a stable pulse at a position midway between the two initial structures [see Fig. 5(b2)]. For longer durations, a doublet of pulses emerges in the rising and falling edges of the initial super-Gaussian seed. An illustration of this process initiated by a 100 ps long seed is provided [see Fig. 5(b3)]. It is worth noting that in this regime, the temporal delay evolves quasi-linearly with the input duration of the weak seed [see dashed line Fig. 5(a)].

In Fig. 5(c), instead of a weak pulsed seed, we considered a weak partially coherent seed. The evolution of this initial random condition demonstrates that pulses tend to emerge on the peak of seeds and near the edge. Generally speaking, when operating in the RR conditions, the temporal intensity gradient of the input coherent or partially coherent seed merely determines the initial positions where the temporal structures tend to develop during the early stages of evolution, while the final timing of stationary pulses is also influenced by the nonlinear dynamics and the gain competition associated with the cavity parameters. From one hand, appropriate gain and wavelength detuning are required to ensure the stability of structures. On the other hand, introducing a controllable seed can realize the manipulation of the pulse distribution. Therefore, the temporal randomness observed in the random pulse train originates from the stochastic nature of noise serving as the seed light during the self-starting process. Once a stable pulse train is established, the Mamyshev SA effectively suppresses the residual noise and the timing of pulse train becomes stable. Under unchanged cavity parameters, each specific temporal structure of the seed light corresponds to a unique stable pulse train. Such stable pulse train follows the aforementioned causality.

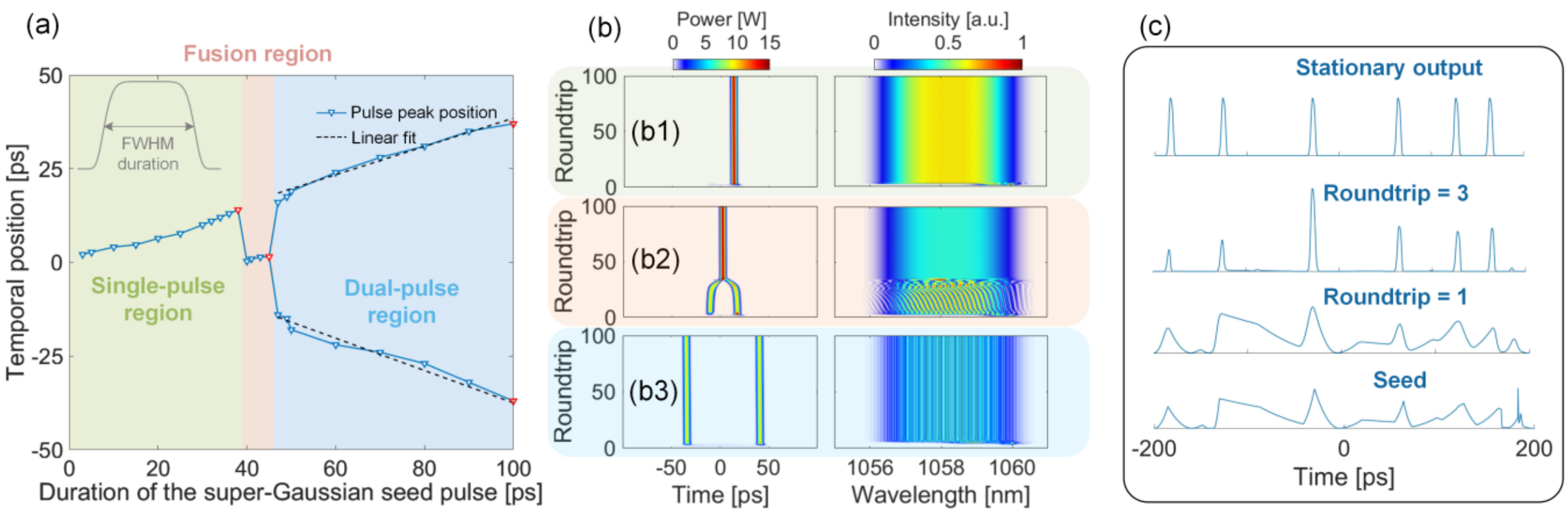

**Fig. 5. Steady-state solution with weak seeds.** (a) Temporal positions of the stable pulse peaks as a function of the FWHM duration of the weak super-Gaussian seed pulse. Three regions are distinguished: dual- pulse region (blue), fusion region (red) and single-pulse region (green). The black dashed lines are linear fits of the trend observed for the leading and trailing pulses in the dual pulse region. (b) Temporal and spectral evolutions of the intensity profiles initiated with a weak super-Gaussian seed with different durations. Top to bottom widths of 38 ps, 45 ps and 100 ps respectively (see red triangles in panel (a)). (c) Temporal evolution of the optical field when using a partially coherent seed. Cavity parameters are $E_{sat}$ = 0.5 nJ and $\delta\lambda$ = 1.8 nm.

In that context, it becomes very attractive to further investigate the use of MOs for bit buffering as it has been explored in coherent cavities or in other fiber laser architectures. [54–56] Indeed, since their first experimental demonstration, MR-based cavities have demonstrated remarkable buffering abilities. [57] In order to validate the potential of the precise temporal manipulation of the intra-cavity pulse train and to get insights on how the cavity reacts following an external high-power pulse seed, [58,59] we conducted simulations on disrupting an established stable pulse train through the external injection of a high-peak-power pulse sequence, as illustrated in Fig. 6. It is worth mentioning that the external seeds are used to manipulate the timing of random pulse train, instead of initiate the mode locking. Specifically, the intra-cavity stabilized pulses are superimposed with the externally injected pulses, forming a completely new pulse train that acts as new conditions for subsequent iterations. When the peak power of the externally injected pulses exceeds 200 W, the original intra-cavity pulse train fails in the gain competition and the final stabilized pulse train becomes synchronized with the externally injected pulse train, which is categorized as "*refresh*". Notably, under the cavity parameter configuration of $E_{sat}$ = 0.5 nJ and $\delta\lambda$ = 1.8 nm, the MO are unable to generate and sustain a stable single pulse, as concluded from the panoramic view. Therefore, we use three high-peak-power pulses as the refresh pulse train [see Fig. 6(c)]. However, the pre-existing pulse train can co-exist with the external injected pulse as long as their intensities are comparable, which is referred as "*writing*" [see Fig. 6(d)]. The composite new conditions force the emergence of pulse train with desired timing, thereby enabling precise timing manipulation of pulse train, named as *timing-injection locking*. In more details, the externally injected pulse train in the simulation consist of identical Gaussian pulses with a pulse width of 10 ps and a central wavelength of $\lambda_0$. A temporal spacing of 50 ps has been chosen in order to avoid the subsequent fusion, ensuring the formation of stable multi-pulse train. Note that, the injected pulse train with lower peak power cannot introduce new stable pulses due to the saturable absorption effect. Figure 6(b) illustrates the ability of continuous manipulation of the pulse train [59] whether refresh or writing at different temporal locations and roundtrips, indicating the potential of the scheme to operate stably over a large range of parameters thanks to the wide range of RR's parameter space.

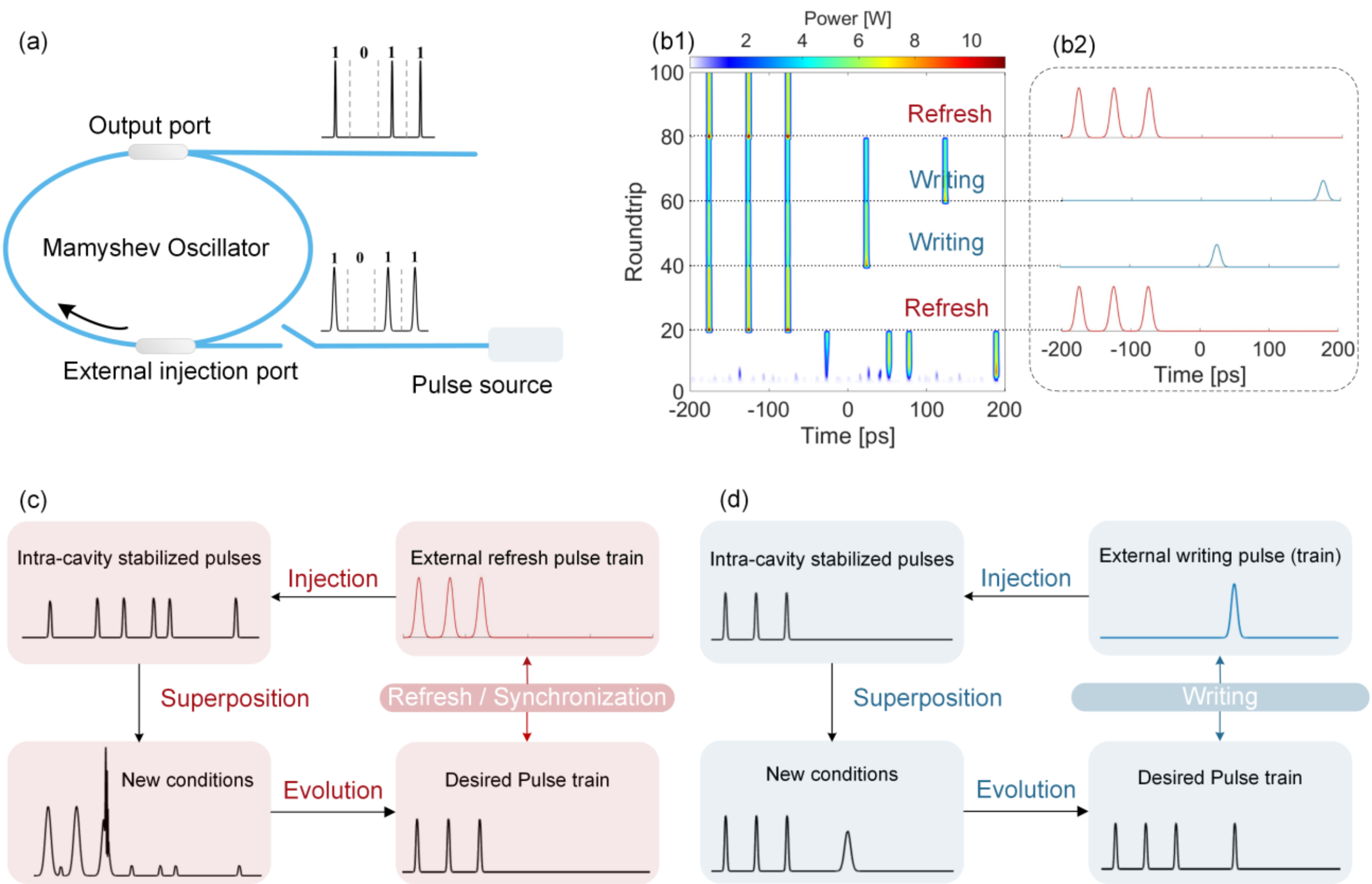

**Fig. 6. Mechanisms of timing-injection locking in a MO.** (a) Schematic diagram of external injection manipulation in a MO. (b1) Temporal evolution of output pulse train over 100 roundtrips. (b2) Timing of external injection pulse train. The refresh pulse trains are injected into the MO at the $20^{th}$ and $80^{th}$ roundtrips. The writing pulse trains are injected into the MO at the $40^{th}$ and the $60^{th}$ roundtrips. (c) Mechanism of synchronization between intracavity pulses and external refresh pulses. (d) Mechanism of writing using writing pulse (train). The cavity parameters are $E_{sat}$ = 0.5 nJ, $\delta\lambda$ = 1.8 nm.

The timing-injection locking scheme unfolds high-capacity and flexible optical storage and buffering of digital information in the MO cavity, which has attracted increasing research interests in recent years. In this numerical study, we also examined the potential on all-optical data storage using this scheme, by which arbitrary arrangement of dissipative solitons can be customized at will. In particular, an eight-bit ASCII code is achieved where '1' and '0' are represented by the existence of a pulse in a time slot, of which the length is limited by the separation avoiding fusion between adjacent dissipative solitons. For instance, 4 letters 'HUST', which is the acronym of the authors' affiliation, are written digitally into the MO in four distinct frames with distance large enough to minimize interactions and ensure the independent manipulation within each frame [see Fig. 7]. Fig. 7(a) demonstrates that the pulse trains representing the letters 'H', 'U', and 'S' are stably maintained in the MO through injecting one single external refresh pulse train modulated by the corresponding signal to perform both refresh and write operations. To demonstrate the re-writing capability of our storage scheme, we encode only the refresh pulse train representing the high four bits [0101] of the letter 'T' at first, and subsequently write the low four bits by injecting the write pulse train [0100] after several roundtrips. There is no doubt that the re-writing scheme requires more precise timing control than the former one.

We should also mention that there are three main factors affecting the number of usable bits in our MO. First, the peak power of the injection pulse train determines the time range that can be encoded in one frame because excessive peak power increases the instability of the laser system. Second, the time interval between injected pulses determines the coding time slot: insufficient time interval results in pulse train deviating from the anticipated outcome due to the fusion of the neighboring pulses. Recent experimental investigation has demonstrated optical buffering functionality in a non-self-starting MO by encoding the seed pulses, but the capacity is limited. [59] Ideally, based on our all-optical storage mechanism, the storage capacity in the MO is given by: $N = \frac{n_g L_t}{c \cdot \delta\tau}$. In this expression, $n_g$ denotes the group refractive index of the fibers, $L_t$ represents the total fiber length, $c$ is the speed of light in vacuum, and $\delta\tau$ refers to the time slot spacing and the minimum value of $\delta\tau$ corresponds to the $f_{max\text{-}gain}$ obtained from the Floquet analysis. To prevent fusions between two adjacent '1' signals, $\delta\tau$ is reasonably set to 50 ps. Therefore, this long cavity configuration could potentially simultaneously support the circulation of more than 10,000 bits, which, to the best of our knowledge, represents the highest optical storage capacity achieved in a mode-locked laser. Note that the storage capacity in the formula is proportional to the cavity length, yet increasing $L_t$ introduces greater pulse broadening due to dispersion effect in the SMFs, which raises the likelihood of fusions. Unfortunately, the storage capacity cannot be significantly enhanced by simply increasing the cavity length in virtue of the accordingly increased time slot $\delta\tau$. The final factor to be taken into account will be the value of $E_{sat}$ and the time of gain recovery that will ultimately limit the number of pulses that can coexist within the cavity.

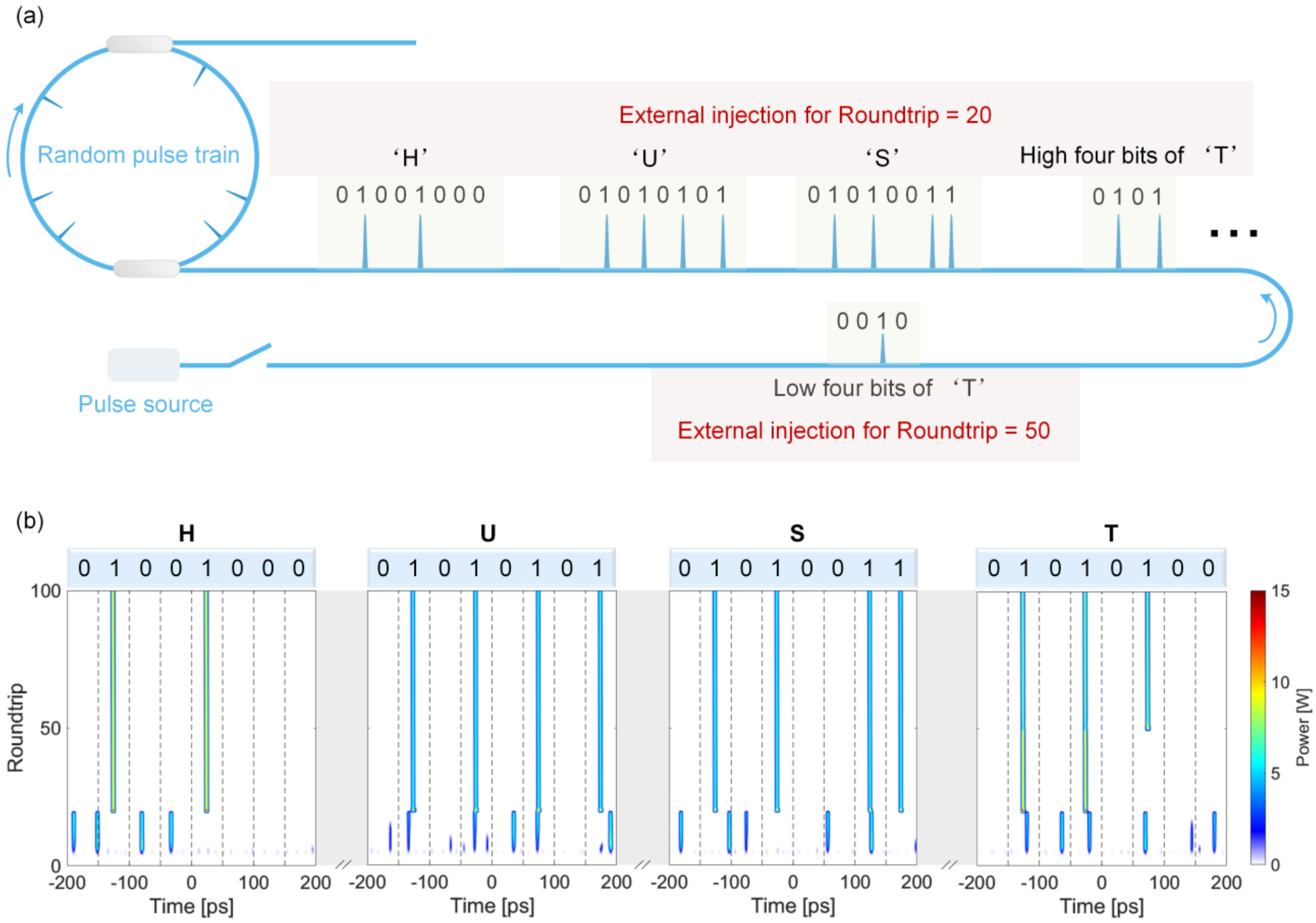

**Fig. 7. All-optical data storage for the acronym of the authors' affiliation (HUST).** (a) Visualization of the storage of the letters 'HUST'. (b) Temporal evolution at the output in different temporal frames. In the first 20 roundtrips, the oscillator generates a random pulse from noise. At the $20^{th}$ roundtrip, the pulse train timing is encoded via external injection. After this point, the oscillator maintains the pulse train with the desired timing without requiring further manipulation in the frames of 'H', 'U' and 'S'. The high 4 bits of the letter 'T' are written at the $20^{th}$ roundtrip by injecting a refresh pulse train, while the low four bits are written at the $50^{th}$ roundtrip by injecting a write pulse train. The key cavity parameters are $E_{sat}$ = 0.5 nJ and $\delta\lambda$ = 1.8 nm.

## 4. Conclusion

In conclusion, we have numerically investigated the DFI-induced self-starting mechanism in an all-normal-dispersion MO,

as well as four operational regimes demonstrating different multi-pulse dynamics under conditions of long cavity and low wavelength detuning. These results highlight a successful solution for the initiation of MOs and the intrinsic property of mode-locked laser systems, providing comprehensive insights into the stability of complex nonlinear systems. Especially in the random operation regime, we reveal the causality of stable pulse distribution and the initial seed, reminiscent of creating on-demand arbitrary pulse sequences. This leads us to detail a novel scheme for programmable generation and storage of light by external injection of high-peak-power pulse trains with designed timing, leading to a new type of all-optical storage system compatible with high-speed and large-capacity optical fiber communication networks. As MRs are able to sustain bidirectional operation [60], MOs can also be used in a bidirectional way [24], therefore doubling the potential capacity of this optical storage.

Building upon these findings and inspired by the time-space duality [61,62], the temporal patterns generated in our cavity could be extended to the spatial domain. Indeed, following the analogy proposed in the ref. [34], where the DFI has been studied from the spatial perspective, one can consider a DFI-based free-space MO as an ideal platform to generate and tailor the spatial solitons or even more complex on-demand patterns. For a broader perspective, the external seed light frames could be duplicated and imaged at the output. Such an all-optical latch based on the MO architecture sheds new ligth on the capturing and imaging for the ultrafast biological or chemistry dyamnics.

## 5. Supporting Information

Supporting information is available from the Wiley Online Library or from the author.

## 6. Acknowledgement

The authors acknowledge funding support from the National Natural Science Funding of China (62275097).

## 7. Conflict of Interest

The authors declare no conflict of interest.

## 8. Data Availability Statement

The data used in this study are available from the corresponding authors under reasonable request.